\documentclass{aa}
\usepackage[bookmarks=false]{hyperref}
\usepackage{epsfig,graphicx,twoopt}
\usepackage{natbib}
\usepackage[varg]{txfonts}
\usepackage{color}
\begin{document}
\title{Simultaneous dual-frequency radio observations of S5~0716+714: A search for intraday variability with the Korean VLBI Network}

\author{Jee Won Lee\inst{1,2}
\and Bong Won Sohn\inst{1,3} 
   \thanks{\emph{Correspondence to:}
     bwsohn@kasi.re.kr}
\and Do-Young Byun\inst{1}
\and Jeong Ae Lee\inst{1,3}
\and Sungsoo S. Kim\inst{2}}

\institute{Korea Astronomy and Space Science Institute 776, Daedeok-daero, Yuseong-gu, Daejeon 34055, Republic of Korea
  \and Department of Astronomy and Space Science, Kyung Hee University, 1732, Deogyeong-daero, Giheung-gu, Yongin-si, Gyeonggi-do 17104, Republic of Korea 
   \and Korea University of Science and Technology, 217, Gajeong-ro, Yuseong-gu, Daejeon 34113, Republic of Korea}
\date{Accepted}
   
\abstract
{This study aims to search for the existence of intraday variability (IDV) of BL Lac object S5 0716+714 at high radio frequencies for which the interstellar scintillation effect is not significant.
Using the 21-meter radio telescope of the Korean VLBI Network (KVN), we present results of multi-epoch simultaneous dual-frequency radio observations.
Single-dish observations of S5 0716+714 were simultaneously conducted at 21.7\,GHz (K-band) and 42.4\,GHz (Q-band), with a high cadence of 30-60 minute intervals.
We observed four epochs between December 2009 and June 2010.
Over the whole set of observation epochs, S5 0716+714 showed significant inter-month variations in flux density at both the K- and Q-bands, with modulation indices of approximately 19\,\% for the K-band and approximately 36\,\% for the Q-band.
In all epochs, no clear intraday variability was detected at either frequency. 
The source shows monotonic flux density increase in epochs 1 and 3 and monotonic flux density decrease in epochs 2 and 4.
In the flux density increasing phases, the flux densities at the Q-band increase more rapidly.
In the decreasing phase, no significant flux density difference is seen at the two frequencies.
The situation could be different close to flux density peaks that we did not witness in our observations.	
We find an inverted spectrum with mean spectral indices, $\bar{\rm {\alpha}}$ ($S_\nu \propto \nu^{-\alpha}$), of -0.57$\pm$0.13 in epoch 1 and -0.15$\pm$0.11 in epoch 3. 
On the other hand, we find relatively steep indices $\bar{\rm {\alpha}}$ of +0.24$\pm$0.14 and +0.17$\pm$0.18 in epochs 2 and 4, respectively.
We conclude that the frequency dependence of the variability and the change of the spectral index are caused by source-intrinsic effects rather than by any extrinsic scintillation effect.
}

\titlerunning{Simultaneous dual-frequency radio observations of S5 0716+714}
\authorrunning{Lee et al.}
\maketitle 

\section{Introduction}
Flux density variability on various time scales in active galactic nuclei (AGNs) has been reported over a broad range of the electromagnetic spectrum.
Intraday variability~(IDV) in AGNs was observed in radio wavelengths as well as in the optical band~\citep[e.g.,][]{1987AJ.....94.1493H, 1990A&A...235L...1W, 1992A&A...258..279Q, 1995ARA&A..33..163W, 2003A&A...401..161K}.
{\citet{1987AJ.....94.1493H} categorized IDVs into two types according to time scales of flux density variability, namely, sources with variability longer than two days as type I and shorter than two days as type II.
IDV has been detected in compact radio sources with flat spectra, such as BL Lacertae (BL Lac) objects and optically violent variable (OVV) quasars.
IDV has been explained as resulting from either superluminal motion of relativistic shock in an inhomogeneous jet~\citep{1985ApJ...298..114M, 1991A&A...241...15Q}, rapid changes of the direction of shocks inside a jet~\citep{1992A&A...259..109G}, or the extrinsic interstellar scintillation (ISS) effect due to scattering in the interstellar plasma screen between the source and the observer~\citep{1989A&A...209..315S, 2002Natur.415...57D}.
The ISS effect is usually dominant at long wavelengths because the scattering effect depends on frequency, according to the relationship $\nu^{-2.2}$~\citep{1984A&A...134..390R}.
Therefore, the ISS effect is not significant at high radio frequency.

S5 0716+714~\citep[$z=0.31\pm0.08$;][]{2008A&A...487L..29N} is a flat-spectrum BL Lac object.
The source is known as a type II IDV source at 2.7\,GHz~\citep{1987AJ.....94.1493H}. 
IDV with a peak-to-peak amplitude of $\sim$20\,\% has also been reported even at 32\,GHz~\citep{2001ASPC..250..184K, 2003A&A...401..161K}.
A close correlation between the radio and the optical band with a short time lag in this source was observed~\citep{1991ApJ...372L..71Q, 1996AJ....111.2187W}.
In multi-frequency campaigns from centimeter to sub-millimeter wavelengths, S5~0716+714 did not exhibit type II IDV, whereas a monotonic increase in the flux density similar to interday variability was detected~\citep{2006A&A...456..117A, 2006A&A...451..797O, 2008A&A...490.1019F}.
\citet{2008A&A...490.1019F} found that the flux density variation is correlated over all observed wavelengths from centimeter to millimeter and that there is a trend of increasing time lag toward lower frequencies.
They also found that the flux density variations tend to be stronger at higher frequencies.
All of these behaviors of the variability were interpreted as evidence of a source-intrinsic origin rather than as part of the ISS effect.
\citet{2012MNRAS.425.1357G} detected the intraday variability with the timescale of less than 1.5~days at the low radio frequencies~(2.7, 4.8, and 10.5~GHz). 
They found more rapid variability (approximately 0.5~days) at 10.5~GHz and that the variability of 10.5~GHz leads the variability of the lower frequencies by approximately 1~day. 
They also found the modulation indices decrease with increasing frequency.
These behaviors suggest that ISS seems to be dominant at 2.7 and 4.8\,GHz, whereas the intrinsic contribution to the source variability predominates at 10.5\,GHz.
\citet{2013A&A...552A..11R} reported that the broadband flux density and variability of S5 0716+714 peak around 43\,GHz that they are saturated above this frequency, and that they are damped at lower frequency due to an opacity effect.
Therefore, observations at 43\,GHz are advantageous for the study of the characteristics of variability in the flux density.

This study was planned to search for type II IDV at the radio frequency and for evidence of intrinsic flux density variability in S5 0716+714. 
In this paper, using the KVN radio telescope, we report on the results of multi-epoch simultaneous dual-frequency observations.


\begin{table}
\caption{Observation dates}
\label{table:obs_date}

\begin{tabular}{ccc}
\hline\hline
\textit{Epoch} &             Date               &   MJD    \\
\hline
1              & 12 Dec 2009 - 15 Dec 2009      & 55177.57 - 55180.92 \\
2              & 5 Jan 2010 - 11 Jan 2010       & 55201.04 - 55207.53 \\
3              & 28 Jan 2010 - 31 Jan 2010      & 55224.80 - 55227.60 \\
4              & 14 Jun 2010 - 16 Jun 2010      & 55361.86 - 55363.71 \\
\hline
\end{tabular}
\end{table}

\begin{table*}
\caption{Descriptions of observations}
\label{table:obs_detail}
\centering
\begin{tabular}{lcc}
\hline\hline
Frequency    & 21.7\,GHz (K-band) & 42.4\,GHz (Q-band)\\
\hline
Bandwidth    & 512\,MHz           & 512\,MHz\\
Half-power beam width (HPBW) & 130$\arcsec$ & 65$\arcsec$\\
System temperature\tablefootmark{*} ($T_{\rm sys}$) & 100, 80, 80, 230\,K & 200, 170, 170, 250\,K \\
Observation mode     &\multicolumn{2}{c}{Cross-Scan in AZ-EL direction} \\
Flux density calibrator     & \multicolumn{2}{c}{3C\,286}\\
Relative gain calibrator     & \multicolumn{2}{c}{S5~0836+710}\\
\hline
\end{tabular}
\tablefoot{
\tablefoottext{*}{System temperatures are written for epoch 1, 2, 3, and 4 in sequence.}}
\end{table*}

\section{Observations and data reduction} \label{obs_reduction}
\subsection{Observations} 
The Korean VLBI Network (KVN), which is a very long baseline interferometry (VLBI) facility for mm-wavelength, allows simultaneous multi-frequency observations~\citep{2004evn..conf..281K}.
KVN consists of three 21-meter Cassegrain radio telescopes that are located in Seoul (Yonsei University), Ulsan (University of Ulsan), and Jeju Island (Jeju International University) in the Republic of Korea \footnote{\url{http://kvn.kasi.re.kr/index.html/main.html}}.

Single-dish observations were performed with the KVN Yonsei radio telescope at four epochs between December 2009 and June 2010 .
The observation dates are listed in Table \ref{table:obs_date}.
In epoch 1, we observed from UT 04:00 to UT 16:00 in 3.5 consecutive days.
In epochs 2 - 4, the observations were conducted over 6.5 days, 3 days, and approximately 2 days, respectively.
However, the observations were ceased due to bad weather conditions for approximately 1 day in epoch 3.
The target and the calibrator were observed every 30 minutes in epochs 1-3 with an on-source integration time of 40 sec at the K-band and 20 sec at the Q-band.
In epoch 4, we observed the sources in every hour with an on-source integration time of 80 sec and 40 sec at the K-band and the Q-band, respectively, due to the high system temperature in summer.
 
Simultaneous dual-frequency observations were conducted at both 21.7\,GHz (K-band) and 42.4\,GHz (Q-band).
The observed bandwidth was 512\,MHz and the system temperatures, $T_{\rm sys}$, were from $\sim$80 to $\sim$230\,K at the K-band and from $\sim$170 to $\sim$250\,K at the Q-band over the whole set of observation epochs.
 
Flux density measurements were performed by cross-scans on the source in the azimuth and elevation directions.
One cross-scan set consists of ten sub-scans in each direction.
The standard deviation of the pointing offsets over the whole set of observation epochs were $\sim$4$\arcsec$ in the azimuth direction and $\sim $8$\arcsec$ in the elevation direction.
Due to a deviation of the beam alignment, the pointing offset between the K-band and the Q-band was $\sim$4$\arcsec$ in the elevation direction.
Pointing offsets larger than 50\,\% of the HPBW (half-power beam width) at the Q-band occurred often in the elevation direction in epoch 1. 
This is because of thermal deflection by the antenna mounting structure.
Therefore, to avoid this effect, we adjusted the antenna pointing during monitoring from epoch 2 to epoch 4.
The primary calibrator, 3C~286, and the secondary calibrator, 0836+710, were observed to convert the antenna temperature to the flux density and to calibrate the time-dependent systematic variations, respectively.
The final errors of the flux densities are 0.16\,Jy at 22\,GHz, 0.42\,Jy at 43\,GHz in epoch 1, and 0.08\,Jy at 22\,GHz and 0.2\,Jy at 43\,GHz in epochs 2-4.
S5 0716+714 and 0836+710 were observed alternately and 3C~286 was observed with a cadence of one or two hours.
To measure the sky opacity at a zenith, $\tau_{0}$, we performed sky-dip observation every one or two hours.
A summary of the observations is provided in Table~\ref{table:obs_detail}.

\subsection{Data reduction}
The data reduction processes for the cross-scan measurements were as follows:
first, we removed the sub-scans that had large fluctuations in power level. 
Those large fluctuations seem to have been affected by atmospheric and instrumental fluctuations. 
We averaged the sub-scans in the azimuth\,(\textit{az}) and elevation\,(\textit{el}) directions separately.
Then, we measured the opacity-corrected antenna temperature $T_{\rm A}^*$, the pointing offset, and the HPBW using a Gaussian profile fit to the average of the sub-scans.
Here, $T_{\rm A}^*$ is defined by $T_{\rm A}^*=T_{\rm A}\cdot \exp^{\tau_{0}sec z}$, where $T_{\rm A}$ is the antenna temperature, $\tau_{0}$ is the opacity at zenith, and $z$ is the angle at zenith.
Then, the corrected pointing offset antenna temperature is 
\begin{equation}
T^*_{{\rm A,corr,}i}=T^*_{{\rm A},i}\cdot \exp\Big(4\cdot \ln 2 \cdot\frac{x^2_{j}}{\theta^2}\Big),   i \neq j
\end{equation}
where, $T^*_{{\rm A},i}$ is the opacity corrected antenna temperature in the \textit{i}-direction and $x_{j}$ indicates the pointing offset in the \textit{j}-direction; $\theta$ is the HPBW (see Table \ref{table:obs_detail})~\citep{2004PhDT, 2013A&A...560A..80P}.
The directions \textit{i} and \textit{j} are the azimuth and the elevation, respectively.
We next determined the corrected antenna temperature $T^*_{\rm A,corr}$, as the arithmetic mean antenna temperature $T^*_{{\rm A,corr},az}$ and $T^*_{{\rm A,corr}, el}$.
For pointing offsets larger than 50\,\% of the HPBW in the elevation direction, only $T^*_{{\rm A,corr},el}$ was taken into account.
In order to remove systematic time-dependent variations (i.e. relative gain) of S5 0716+714, we assumed that $T^*_{\rm A,corr}$ of 0836+710 is a constant, $\bar{T}_{{\rm A},\rm corr}^*$, over a few days; we then computed the relative gain using $\bar{T}_{{\rm A},\rm corr}^*$/$T_{{\rm A},\rm corr}^*$ of 0836+710.
The deviations of the relative gain are $\leq$4\,\% at the K-band and $\leq$7\,\% at the Q-band. 
These values correspond to a 1-$\sigma$ measurement error.
We linearly interpolated the relative gain of 0836+710 with time and determined a relative gain value with time of 0716+714.
$T^*_{{\rm A,corr},gain}$ of S5 0716+714 was obtained by multiplying the relative gain and $T^*_{\rm A,corr}$ of S5 0716+714.  
Finally, $T^*_{{\rm A,corr},gain}$ was converted to flux density using conversion factors obtained from the primary calibrator, 3C~286.
Here, the flux density values of 3C~286 were 2.64\,Jy at the K-band and 1.5\,Jy at the Q-band; these values were obtained from measurements of Mars (Sohn et al. in prep.). 
Errors in the calibrated flux density of S5 0716+714 were calculated by propagating the measurement uncertainties of both the source and the calibrator. 

\begin{figure}
\centering
  \includegraphics[width=9cm]{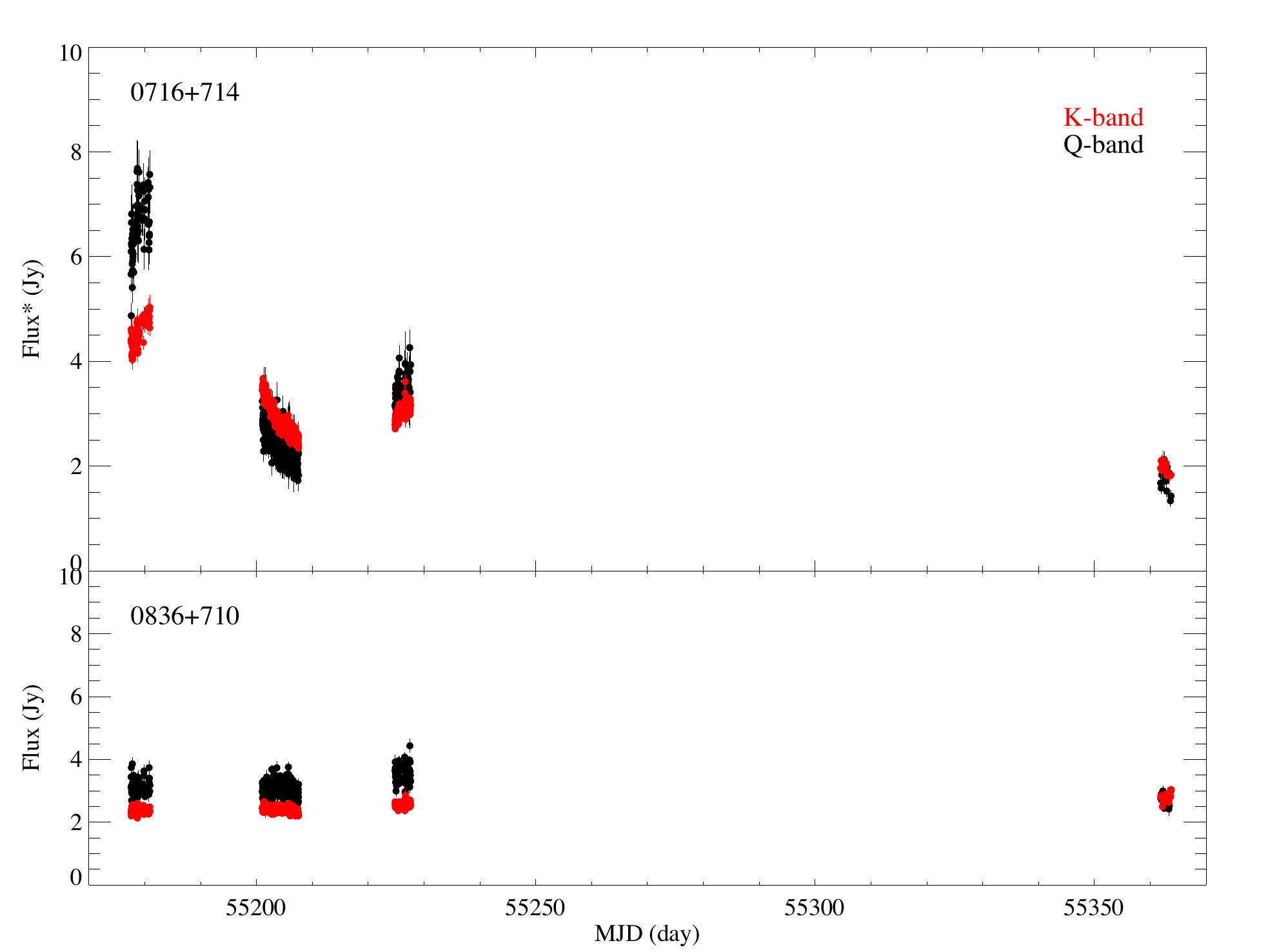}
    \caption{Light curves of S5~0716+714 (top panel) and of the secondary calibrator, 0836+710, (bottom panel) over the whole set of observation epochs. Red and black symbols indicate the K-band and the Q-band, respectively. Here, the error bars correspond to $1\sigma$ measurement error.}
    \label{fig:lc_total}
\end{figure}

\section{Results}\label{result}
\subsection{Light curves}\label{result:light curve} 
The light curves of S5 0716+714 (top) and 0836+710 (bottom), obtained at the K-band (red symbols) and the Q-band (black symbols) are shown in Fig.~\ref{fig:lc_total}.
The light curves include data over the whole set of observation epochs.
As can be seen in Fig.~\ref{fig:lc_total}, S5 0716+714 exhibits significant flux density variation at both frequencies.
During our observations, the source was brightest in epoch~1, with mean flux densities of 4.5$\pm$0.3\,Jy and 6.6$\pm$0.6\,Jy at the K- and the Q-bands, respectively.
In epoch 2, the mean flux densities were 2.9$\pm$0.3\,Jy and 2.5$\pm$0.3\,Jy and, in epoch 3, the mean flux densities were 3.1$\pm$0.2\,Jy and 3.4$\pm$0.3\,Jy at the K-band and the Q-band, respectively.
In epoch 4, the source was faintest, with mean flux densities of 1.9$\pm$0.1\,Jy at the K-band and 1.8$\pm$0.2\,Jy at the Q-band.

\begin{figure}
  \includegraphics[width=9cm]{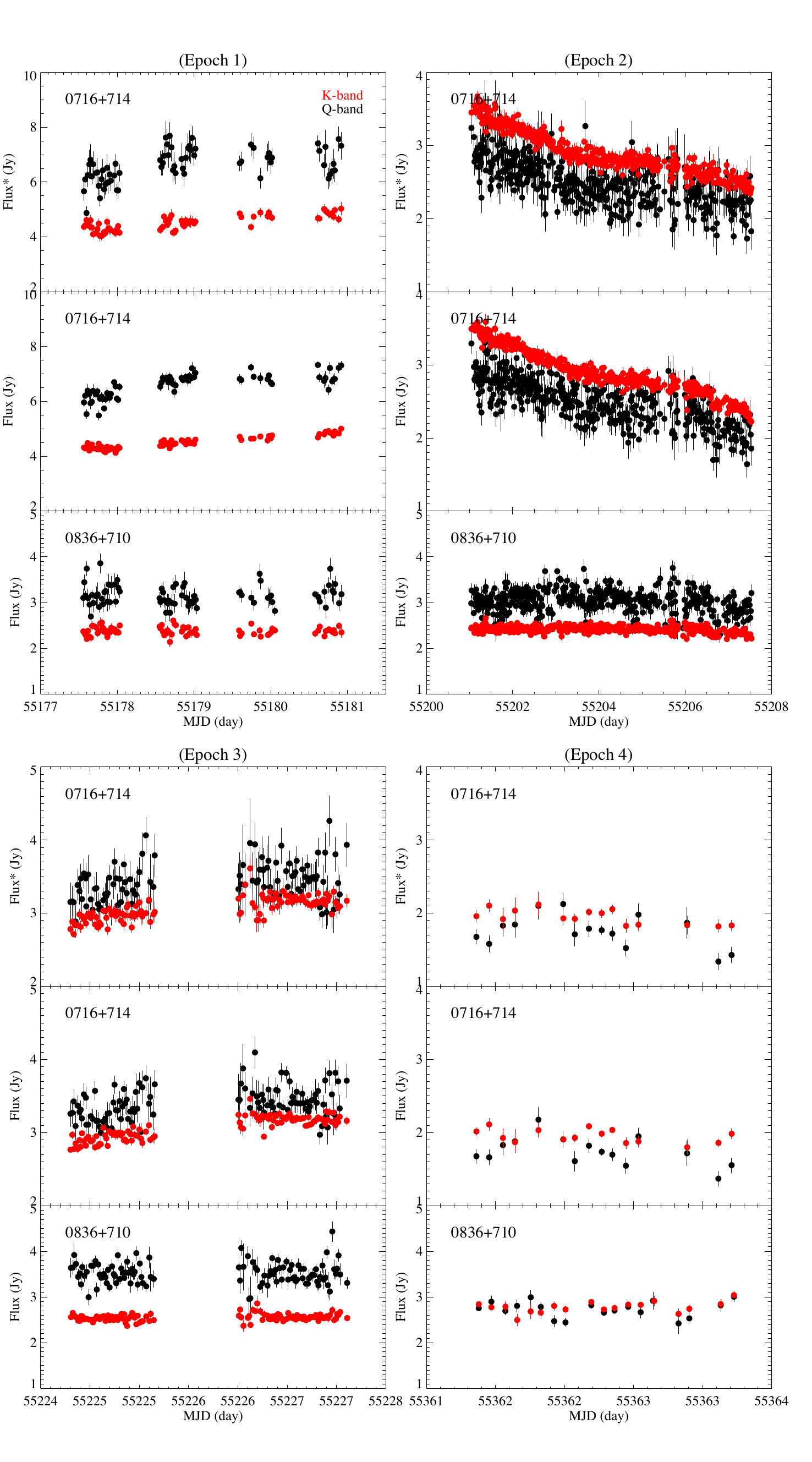}
    \caption{Light curves of S5~0716+714 (top and middle panels) and of the secondary calibrator, 0836+710, (bottom panel) obtained from multi-epoch observations at the K-band and the Q-band simultaneously. In the middle panel of each epoch, the flux densities of S5 0716+714 correspond to the flux densities without the calibration, with relative gain of 0836+710. The top panel represents the finally-calibrated flux densities of S5 0716+714. Red and black symbols indicate the K-band and the Q-band, respectively. Here, the error bars indicate $1\sigma$ measurement error.}
    \label{fig:lc_all}
\end{figure}

Figure~\ref{fig:lc_all} shows light curves of the source for each epoch, separately.
In each epoch, it seems that there is fast flux density flickering within a one-day time scale.
However, the amplitudes are smaller than the 2-$\sigma$ measurement error, and thus we considered that those variations are insignificant.
In our multi-epoch observations, we detected two different trends in the flux density of S5 0716+714.
The source clearly shows a monotonic increase in the flux density in epochs 1 and 3 and a monotonic flux density decrease in epochs 2 and 4 at both frequencies.
In epoch 2, the light curves show changes of slope.
In epochs 1 and 3, the flux density at the Q-band is brighter than that at the K-band, whereas in epochs 2 and 4, the flux density at the Q-band is dimmer than that at the K-band.
To search for possible time lags, we performed cross-correlation analysis on the light curves for four epochs. 
However, we found no intraday-scale time lag between the K-band and the Q-band in four epochs.
This is mainly because the flux density variations are linearly increasing or decreasing within the intraday time scale. 
Interestingly, the IDV light curves at 2.7, 4.8, and 10.5\,GHz of \citet{2012MNRAS.425.1357G}, whose observations overlap with our epoch 1, show different trends.
The flux density of the source decreases at low frequencies whereas it increases at K-band and Q-band.
This difference can be explained if the variation of flux displays a time lag of up to a few days (depending on frequency), such that the variations in flux density at the higher frequency start ahead of those at the lower frequency, as reported by \citet{2013A&A...552A..11R}.
In this context, the flux density increase at 10.5\,GHz from JD 2455179.5~\cite{2012MNRAS.425.1357G} could be interpreted as lagged behind the flux density increase at 22 and 43\,GHz which began a few days ahead.

\subsection{Variability} \label{result:variability}

To quantify the variability of S5~0716+714 at the K- and the Q-bands for each epoch, we followed statistical variability analysis methods such as modulation index $m$, variability amplitude $Y$, and reduced $\chi^2$-test, as defined by \citet{1987AJ.....94.1493H} and \citet{2003A&A...401..161K}.
To characterize the variability strength of the source, the modulation index $m$ and the variability amplitude $Y$ were calculated. 
The modulation index $m$ is defined as
\begin{equation}
m\,[\%]=100\cdot\frac{\sigma_{S}}{\langle S \rangle},
\end{equation}
where $\sigma_{S}$ and $\langle S \rangle$ denote the standard deviation of the flux density and the mean flux density, respectively.
To take into account the residual variability due to calibration error of the calibrator source, the variability amplitude $Y$ was used with
\begin{equation}
Y\,[\%]=3\sqrt{m^{2}-m^{2}_{0}},
\end{equation}
where $m_{0}$ is a modulation index of 0836+710.
To prove the presence of the variability in the light curves, we performed a reduced $\chi^2$-test with a hypothesis of a constant model as
\begin{equation}
\chi_{\rm r}^2=\frac{1}{N-1}\sum_{i=1}^{N}(\frac{S_{i}-\langle S \rangle}{\Delta{S}_{i}})^2,
\end{equation}
where $S_{i}$ is the individual flux density at time \textit{i}, $\langle$S$\rangle$ is the average of the flux density, $\Delta{S}_{i}$ is the individual measurement error, and \textit{N} is the number of data points.
To reject a constant model in which the source has no variability, we chose a $p$-value cut-off of 0.001.
This cut-off corresponds to 99.9\,\% significance level for variability.

\begin{table*}
\caption{Statistical results for S5~0716+714 and 0836+710. The columns, from first to last, denote epoch, source, observing frequency $\nu$, total number of measurements N, mean flux density <s>, standard deviation of flux density $\sigma_{s}$, modulation index m, variability amplitude Y, reduced $\chi^2$ $\chi^2_{\rm r}$, and corresponding value of the 99.9\,\% significance level of variability $\chi^2_{99.9\%}$.}
\label{table:stat}
\centering
\begin{tabular}{cccccccccccc}
\hline\hline
Epoch  & Source   & $\nu$ & N   & <s>  & $\sigma_{s}$  & m     &  Y     &  $\chi^2_{\rm r}$  & $\chi^2_{99.9\%}$  \\
       &          & (band)&     & (Jy) & (Jy)          & (\%)  & (\%)   &                    &        \\          
\hline
1      & 0716+714 & K    & 67  & 4.5 & 0.3             & 6.7   & 15.7   & 2.96               & 1.63 \\
       &          & Q    & 67  & 6.6 & 0.6             & 9.1   & 19.7   & 2.81               & 1.63  \\
       & 0836+710 & K    & 74  & 2.4 & 0.1             & 4.2   & --     & 1.73               & 1.59  \\
       &          & Q    & 74  & 3.2 & 0.2             & 6.3   & --     & 1.65               & 1.59  \\
\hline
2      & 0716+714 & K    & 411 & 2.9 & 0.3             & 10.3  & 28.2   & 12.96              & 1.23\\ 
       &          & Q    & 411 & 2.5 & 0.3             & 12.0  & 29.9   & 2.27               & 1.23\\
       & 0836+710 & K    & 411 & 2.4 & 0.1             & 4.2   & --     & 1.83               & 1.23\\
       &          & Q    & 411 & 3.0 & 0.2             & 6.7   & --     & 2.45               & 1.23\\
\hline
3      & 0716+714 & K    & 110 & 3.1 & 0.2             & 6.5   & 15.8   & 3.69               & 1.47\\
       &          & Q    & 110 & 3.4 & 0.3             & 8.8   & 20.1   & 1.32               & 1.47\\
       & 0836+710 & K    & 107 & 2.6 & 0.1             & 3.8   & --     & 1.69               & 1.48\\
       &          & Q    & 107 & 3.5 & 0.2             & 5.7   & --     & 2.16               & 1.48\\
\hline
4      & 0716+714 & K    &  15 & 1.9 & 0.1             & 5.3   & 11.7    & 1.28               & 2.58\\
       &          & Q    &  15 & 1.8 & 0.2             & 11.1  & 24.8   & 2.84               & 2.58\\
       & 0836+710 & K    &  18 & 2.8 & 0.1             & 3.6   & --     & 2.74               & 2.40\\
       &          & Q    &  18 & 2.7 & 0.2             & 7.4   & --     & 1.98               & 2.40\\ 
\hline\hline
Whole  & 0716+714 & K    & 603 & 3.1 & 0.6             & 19.4  & 57.1   & 27.80              & 1.19\\
epochs &          & Q    & 603 & 3.6 & 1.3             & 36.1  & 104.3  & 21.28              & 1.19\\
       & 0836+710 & K    & 610 & 2.6 & 0.1             & 3.8   & --     & 3.83               & 1.19\\
       &          & Q    & 610 & 3.1 & 0.3             & 9.7   & --     & 4.32               & 1.19\\       
\hline
\end{tabular}
\end{table*}

We applied the above statistical analysis methods to both S5 0716+714 and 0836+710 at the K-band and the Q-band, respectively. 
The results of the analysis of the modulation index $m$, the variability amplitude $Y$, the reduced $\chi^2$-test, and reduced $\chi^2$ value corresponding to a 99.9\,\% significance level of variability are listed in Table\,\ref{table:stat}.
We finnd significant inter-month flux density variations and frequency dependence of the variability in S5 0716+714, with $m$ values of approximately 19\,\% at the K-band and approximately 36\,\% at the Q-band over the whole set of observation epochs.
These values are larger than those of $m_{0}$ of the secondary calibrator, 0836+710, by factors of four to five at both frequencies.
The frequency dependence of $m_{0}$ in each epoch is thought to be due to atmospheric effect and residual calibration errors.
Although the values of $m$ for 0836+710 are somewhat high, with ranges of 3.6\,\% to 4.2\,\% at the K-band and 5.7\,\% to 7.4\,\% at the Q-band in each epoch, we find larger amplitudes of the flux density variation with the values of $m$ of S5 0716+714 in the ranges of 6.5\,\% to 10.3\,\% at the K-band and 8.8\,\% to 12\,\% at the Q-band in each epoch.    
These variations seem to indicate the monotonic increase or decrease in the flux density of S5 0716+714, rather than indicating type II intraday variability.
We note a higher variability amplitude Y of the source at the higher frequency. 
\citet{2008A&A...490.1019F} found similar frequency dependence of the variability from cm to sub-mm data and explained it as a source-intrinsic origin. 
In epoch 2, the highest Y values seem to be due to the longest duration of the observation period.

To measure a monotonically increasing and decreasing rate in the flux density, we applied linear least-square fitting to the light curves.
Because the light curves in epoch 2 have two different slopes, we divided the light curves into two parts from MJD 2455201.04 to JD 2455203.97 and from JD 2455203.98 to JD 2455207.53.
For epoch 3, we did not conduct the linear least-square fitting to the light curves because of the absence of data for the one day between MJD 2455225.7 and 2455226.5 and the large scatter around MJD 5227.5 that seems to be affected by the scattering of the secondary calibrator, 0836+710.   
In Table\,\ref{table:increase_rate}, we summarize the results of the linear least-square fitting. 
The results show that when the flux density increases in epoch 1, the increase of the rate at the Q-band is faster than at the K-band by a factor of 2.
When the flux density decreases in epoch 2, the decrease of the rate is not very different at the two frequencies.
This behavior of the variability could be different close to other flux density peaks that we did not witness in our observations.
We will discuss this question in our next paper, which is in preparation.

\begin{figure}
\centering
  \includegraphics[width=8cm]{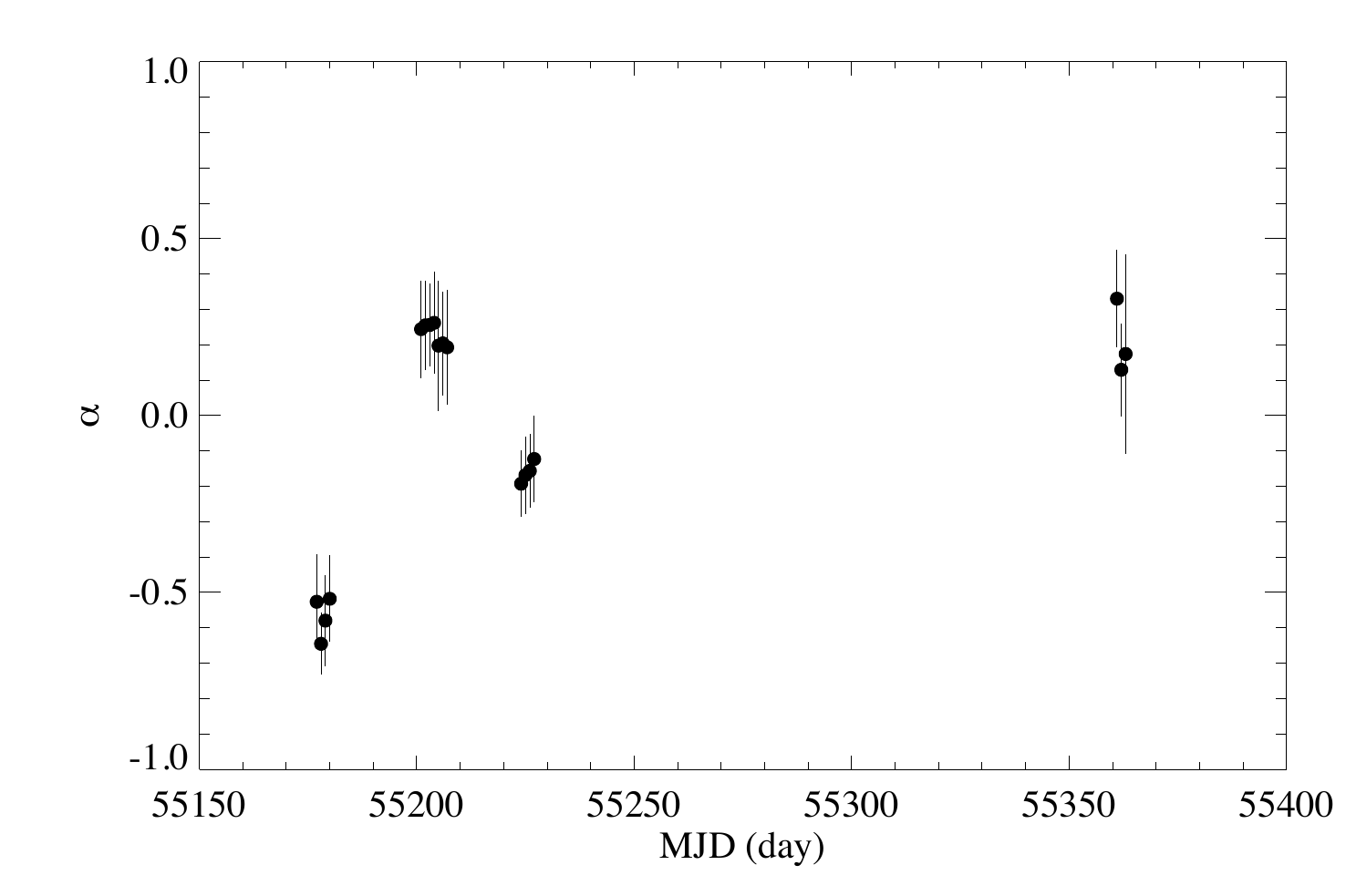}
   \caption{Daily average spectral index of S5 0716+714 over the whole set of observation epochs.}
    \label{fig:s_index}
\end{figure}

\subsection{Spectral index} \label{result:spectral index}
The variation of the spectral index with time, $\alpha$~(defined by $S_{\nu} \propto \nu^{-\alpha}$, where $S_{\nu}$ is the flux density at the observing frequency, $\nu$), of S5~0716+714 is shown in Fig.\,\ref{fig:s_index}. 
In this figure, the data points indicate the daily mean spectral indices $\bar{\rm {\alpha}}$.
The spectral indices vary significantly over the whole set of the observation epochs, with changes in sign from negative to positive.
There are values of a $\bar {\rm {\alpha}}$ of -0.57$\pm$0.13 in epoch 1, +0.24$\pm$0.14 in epoch 2, -0.15$\pm$0.11 in epoch 3, and +0.7$\pm$0.18 in epoch 4.
Interestingly, during epochs of increasing flux density (epochs 1 and 3), the spectra appear inverted (optically thick), whereas while the flux density is decreasing (epochs 2 and 4) the emission appears slightly steep (optically thin).
These spectral changes with observation epoch can be understood as showing that the flux density variability of the source has to be considered as intrinsic in origin. 

To investigate the changes of the spectral indices with the intensity of the flux densities, we compared the spectral indices and the flux densities over the whole time range of our observations and the results are plotted in Fig.\,\ref{fig:flux_s_index}.
This figure shows a tight correlation between the flux densities and the spectral indices.
We estimated the Pearson correlation coefficient $r$ between the flux densities and the spectral indices as shown in Fig.\,\ref{fig:flux_s_index}.
The correlations are strong (i.e., $|r|$ > 0.5) and negative at both frequencies.
This means that when the source becomes bright, the spectrum between the K-band and the Q-band is inverted, whereas when the source becomes fainter, the spectrum is mildly steep.
We suggested that the variable component may emerge from the inner part of a jet and propagate outward from the jet base.

\begin{figure}
\centering
  \includegraphics[width=8cm]{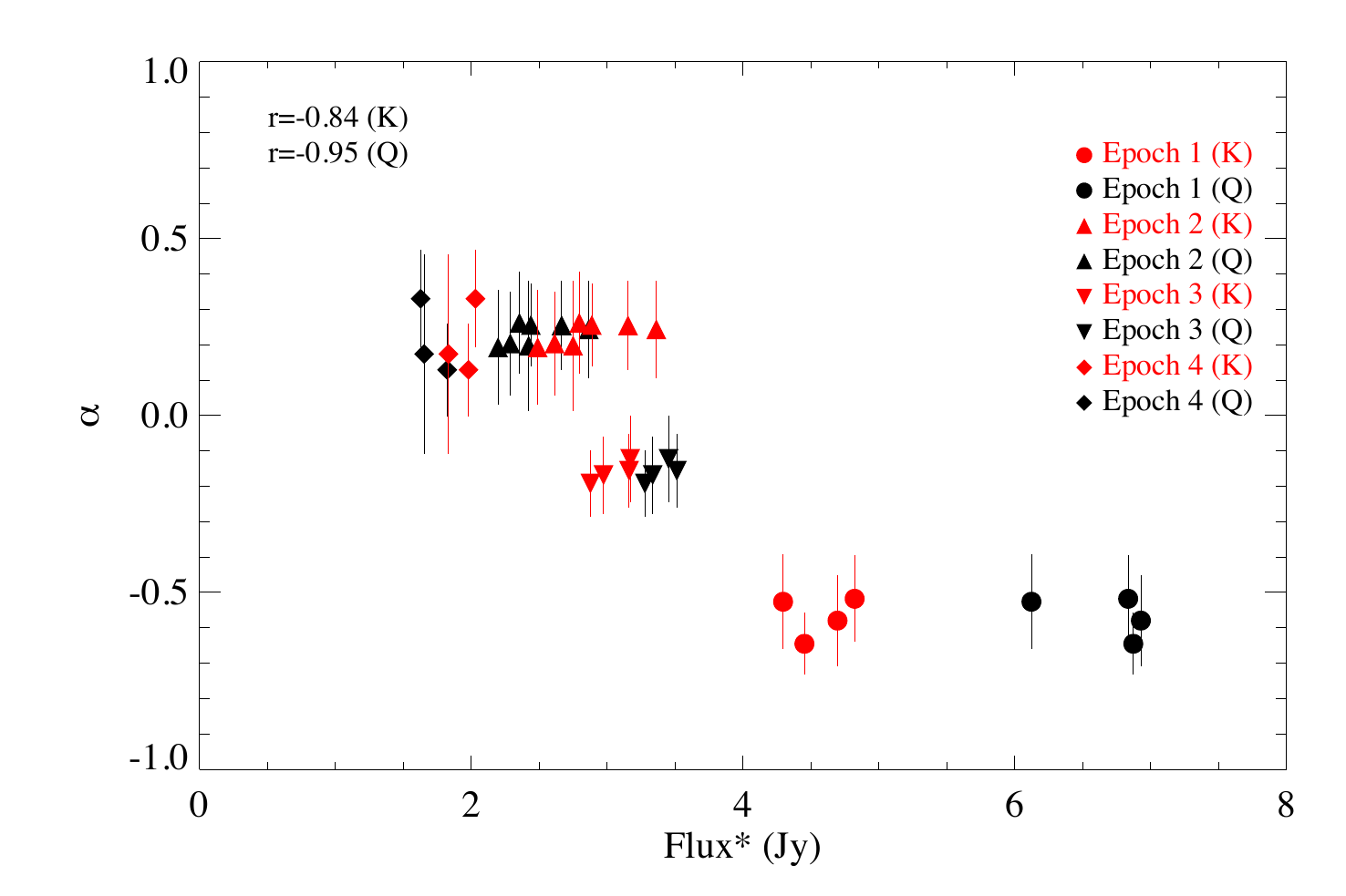}
   \caption{Flux densities versus daily average spectral indices of S5 0716+714 for all four epochs. Red and black symbols correspond to the K-band and the Q-band, respectively. Circle, upward triangle, downward triangle, and diamond symbols indicate epochs 1, 2, 3, and 4, respectively.}
    \label{fig:flux_s_index}
\end{figure}

\begin{table}
\caption{Results of linear least-square fitting to the light curves at the K- and Q-bands.}  
\label{table:increase_rate}
\begin{tabular}{ccccc}
\hline\hline
      &\multicolumn{2}{c}{K-band} & \multicolumn{2}{c}{Q-band} \\
\cline{2-3}\cline{4-5}
Epoch                   & Slope          & $\chi^2_{\rm r}$ & Slope           & $\chi^2_{\rm r}$    \\
                        & (Jy/day)       &                  & (Jy/day)        &                \\
\hline

1                      & 0.18$\pm$0.02  & 1.2               & 0.34$\pm$0.05   &  1.7   \\
\hline
2-1 \tablefootmark{*}  & -0.25$\pm$0.01 & 0.8               & -0.20$\pm$0.02  & 1.1    \\
2-2 \tablefootmark{+}  & -0.11$\pm$0.01 & 1.0               & -0.06$\pm$0.01  & 1.3     \\               
\hline
4                      & -0.11$\pm$0.04 & 0.8              & -0.15$\pm$0.06   & 2.2     \\                 
\hline
\end{tabular}    
\tablefoot{
\tablefoottext{*}{J.D. 2455201.04-2455203.97}  
\tablefoottext{+}{J.D. 2455203.98-2455207.53}
}         
\end{table}
\section{Conclusions}\label{conclusion}
In this work, we search for the existence of type II IDV in the flux density of BL Lac object S5 0716+714 at radio frequencies.
We perform multi-epoch simultaneous dual-frequency observations at the K- and the Q-bands, using the KVN Yonsei radio telescope.
In conclusion, the source shows significant inter-month variation in the flux density at the K- and the Q-bands with a large modulation index over the whole set of observation epochs. 
Despite several intensive observations, no typical type II IDV was found in either frequency in any of the epochs.
The source exhibits a monotonic flux density increase or decrease in each epoch, with increasing variability amplitudes at high frequency.
In the flux density increasing phase, the flux density varies more rapidly at the Q-band whereas in the decreasing phase, there are no significantly different rates for the two frequencies.
We observed variations of the spectral indices over the whole set of observation epochs, with changes in sign from negative to positive. 
We suggest that this variability behavior could have an intrinsic origin rather than resulting from the extrinsic scintillation effect.
In our observations, we did not find statistically meaningful IDV phenomena at 22 and 43\,GHz.
To understand the origin of variability on inter-month time scales, continuous flux density monitoring of S5 0716+714 will be required.

\begin{acknowledgements}
We would like to thank the anonymous referee for important comments and suggestions that have improved the manuscript.
This work was supported by the KASI-Yonsei DRC program of the Korea Research Council of Fundamental Science and Technology (DRC-12-2-KASI) and the International Research \& Development Program of the National Research Foundation of Korea (NRF) funded by the Ministry of Science, ICT and Future Planning (MSIP) of Korea (Grant number: NRF-2012K1A3A7A03049606).
\end{acknowledgements}

\bibliographystyle{aa}  
\bibliography{reference_s1} 
\clearpage

\end{document}